\documentclass[runningheads]{llncs}
   
\usepackage{tikz}
\usepackage{comment}
\usepackage{amsmath,amssymb} 
\usepackage{color}
\usepackage{orcidlink}
\usepackage{subcaption}
\usepackage[labelformat=simple]{subcaption}

\usepackage{multirow}
\usepackage[accsupp]{axessibility}  

\usepackage{hyperref}  
\hypersetup{
     colorlinks=true,
     breaklinks=true,
     linkcolor=black,
     citecolor=black,
     filecolor=black,
     urlcolor=black,
 } 

\usepackage{graphicx}
\usepackage{amsmath}
\usepackage{amssymb}
\usepackage{booktabs}

\usepackage{float} 
\usepackage{mathtools}
\usepackage{url}
\usepackage{tabularx}
\usepackage{footmisc}
\usepackage{balance}
\usepackage{array}
\usepackage{colortbl}
\usepackage{caption}
\usepackage{cite}

\usepackage{cleveref}
\crefname{table}{Tab.}{Tab.}
\crefname{figure}{Fig.}{Fig.}

\crefname{equation}{Eq.}{Eq.} 

\def\ie{\textit{i.e.}~}
\def\wrt{\textit{w.r.t.}~}
\def\eg{\textit{e.g.}~}

\newcommand{\xredit}[1]{\textcolor{black}{#1}}

\newcommand{\repeatthanks}{\textsuperscript{\thefootnote}}

\begin{document}
\pagestyle{headings}
\mainmatter
\def\ECCVSubNumber{4413}  

\title{Lightweight Attentional Feature Fusion: A New Baseline for Text-to-Video Retrieval}

\titlerunning{LAFF for Text-to-Video Retrieval}
%
\author{Fan Hu\inst{1,2}\thanks{Equal contribution.} \and
Aozhu Chen\inst{1,2}\repeatthanks \and
Ziyue Wang\inst{1,2}\repeatthanks \and Fangming Zhou\inst{1,2} \and Jianfeng Dong\inst{3} \and Xirong Li\inst{1,2}\thanks{Corresponding author: Xirong Li (xirong@ruc.edu.cn)}\orcidlink{0000-0002-0220-8310} }
\authorrunning{F. Hu et al.}
%
\institute{MoE Key Lab of DEKE, Renmin University of China \and
AIMC Lab, School of Information, Renmin University of China
\and
College of Computer and Information Engineering, Zhejiang Gongshang University
}
\maketitle

\begin{abstract}
In this paper we revisit \emph{feature fusion}, an old-fashioned topic, in the new context of text-to-video retrieval. Different from previous research that considers feature fusion only at one end, let it be video or text, we aim for feature fusion for both ends within a unified framework. We hypothesize that optimizing the convex combination of the features is preferred to modeling their correlations by computationally heavy multi-head self attention. We propose Lightweight Attentional Feature Fusion (LAFF). LAFF performs feature fusion at both early and late stages and at both video and text ends, making it a powerful method for exploiting diverse (off-the-shelf) features. The interpretability of LAFF can be used for feature selection. Extensive experiments on five public benchmark sets (MSR-VTT, MSVD, TGIF, VATEX and TRECVID AVS 2016-2020) justify LAFF as a new baseline for text-to-video retrieval. 

\keywords{Text-to-video retrieval, video/text feature fusion}
\end{abstract}

\section{Introduction}
Text-to-video retrieval is to retrieve videos \wrt to an ad-hoc textual query from many \emph{unlabeled} videos. Both video and text have to be embedded into one or more cross-modal common spaces for  text-to-video matching. The state-of-the-art tackles the task in different approaches, including novel networks for query representation learning \cite{wray2019fine_r1,sigir20-yang-vr}, multi-modal Transformers for video representation learning \cite{gabeur2020multi,bain2021frozen_r6}, hybrid space learning for interpretable cross-modal matching \cite{Dong2021DE_hybrid,WuN2020}, and more recently CLIP2Video \cite{fang2021clip2video} that learns text and video representations in an end-to-end manner. Differently, we look into \emph{feature fusion}, an important yet largely underexplored topic for text-to-video retrieval.

Given video/text samples represented by diverse features, feature fusion aims to answer a basic research question of \emph{what is the optimal way to combine these features?} By optimal we mean the fusion shall maximize the retrieval performance. Meanwhile, the fusion process shall be explainable to interpret the importance of the individual features. As the use of each feature introduces extra computational and storage overheads, the explainability is crucial for the fusion process to be selective to balance the performance and the cost.

Feature fusion is not new by itself. In fact, the topic has been extensively studied in varied contexts such as \xredit{multimedia content analysis \cite{mmsurvey-2005,mmsurvey-2010} and multimodal or multi-view image classification \cite{aaai17-multi-view,mmsurvey-2018}}. These earlier efforts focus on combining hand-crafted features, because such kinds of features are known to be domain-specific, suffering from the semantic gap problem \cite{pami00-cbir}, and thus insufficient for content representation when used alone. While current deep learning features are already more powerful than their predecessors, no single feature appears to rule all. Dark knowledge about objects and scenes is better carried in pre-trained 2D convolutional neural networks (2D-CNNs) \cite{2021clip_icml}, while 3D-CNNs are more suited for representing actions and motions \cite{ircsn}. For text-to-video retrieval, there are some initial efforts on combining diverse deep video features, \eg JE \cite{mithun2018learning_r2,mithun2019joint_r3}, CE~\cite{liu2019use} and MMT \cite{gabeur2020multi}, whilst W2VV++ \cite{LiXirong2019W2VVPP} and SEA \cite{LiXirong2020SEA} show the potential of combining different text features for better query representation. The recent CLIP series \cite{luo2021clip4clip,fang2021clip2video}, due to their end-to-end learning paradigm, actually lacks the ability of exploiting existing features. Therefore, even in the era of deep learning, the need for feature fusion remains strong.

Concerning approaches to feature fusion, vector concatenation is commonly used when combining features at an early stage \cite{LiXirong2019W2VVPP,Dong2021DE_hybrid}. As for late fusion, multiple feature-specific common spaces are learned in parallel, with the resultant similarities combined either by averaging \cite{mithun2018learning_r2,LiXirong2020SEA}, \xredit{empirical weighting \cite{mithun2019joint_r3}} or by Mixture of Experts (MoE) ensembles \cite{liu2019use}. As the number of features grows, vector concatenation suffers from the curse of dimensionality, while constructing common spaces per feature lacks inter-feature interactions. Moreover, the prior works focus either on the video end or on the text end.
To the best of our knowledge, no attempt is made to develop a unified learning-based approach that works for both ends in the context of text-to-video retrieval, see \cref{tab:related_work}.

One might consider feature fusion by Multi-head Self-Attention (MHSA), the cornerstone of Transformers \cite{vaswani2017attention}. As \cref{fig:mhsa} shows, MHSA transforms a specific feature by blending it with information from all other features, with the blending weights produced by a self-attention mechanism termed QKV. Note that the module was initially developed for NLP tasks, for which exploiting element-wise correlations is crucial for resolving semantic ambiguity. However, as video features extracted by distinct 2D-CNNs and 3D-CNNs are meant for describing the video content from different aspects, we conjecture that optimizing their combination is preferred to modeling their correlations. \xredit{Moreover, the self-attention in MHSA, computed by $\mbox{Softmax}((\frac{QK^{T}}{\sqrt{d_{v}}})V)$, depends largely on inter-feature correlations. It thus tends to have a group effect that features related to each other will be more attended. Consequently, the related yet relatively weak features will be over-emphasized.
Hence, despite its high prevalence in varied contexts, we consider MHSA suboptimal for the current task.}

\newlength{\twosubht}
\newsavebox{\twosubbox}

\begin{figure}[htp]

\sbox\twosubbox{%
  \resizebox{\dimexpr\textwidth-1em}{!}{%
    \includegraphics[height=3cm]{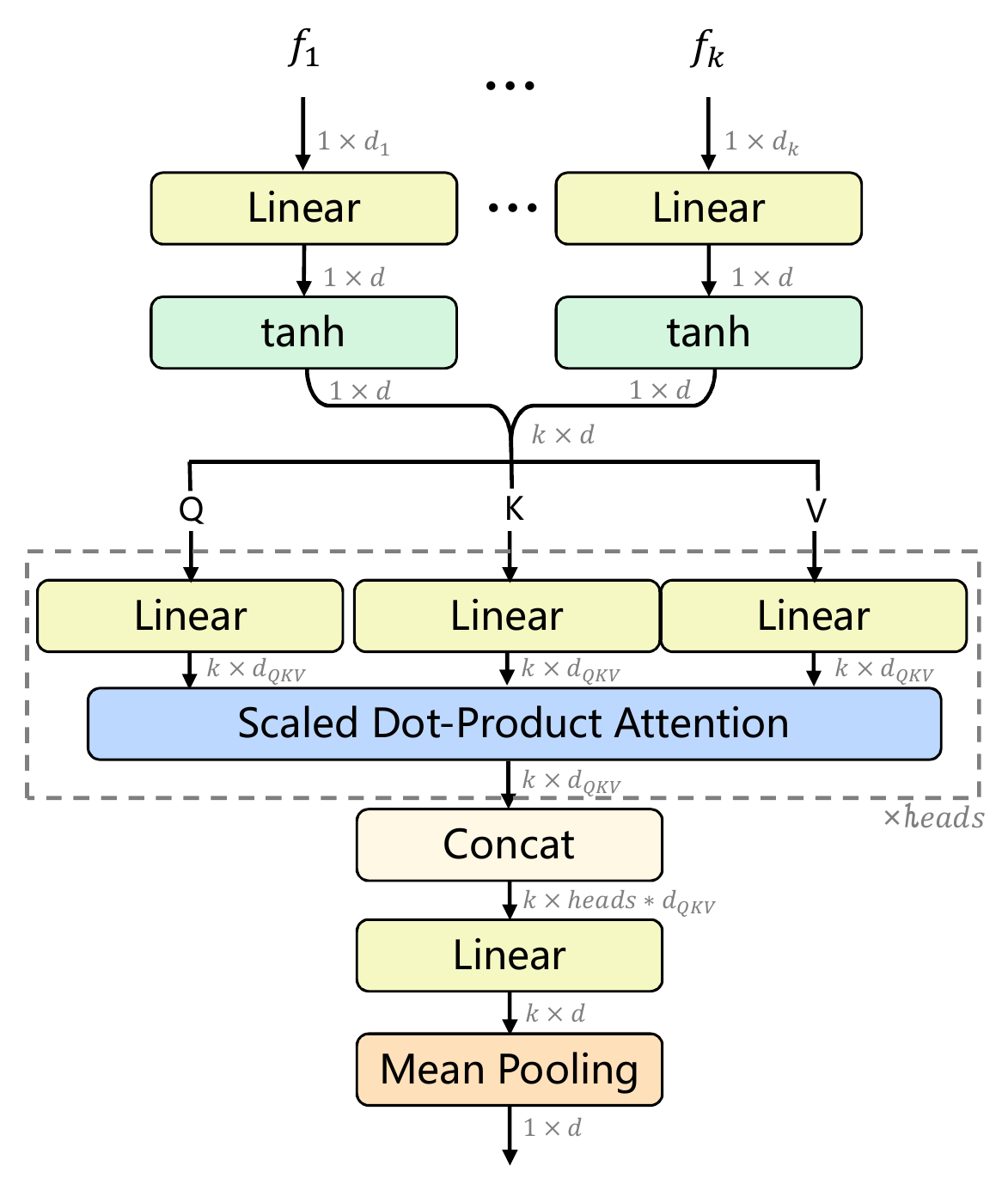} 
    \includegraphics[height=3cm]{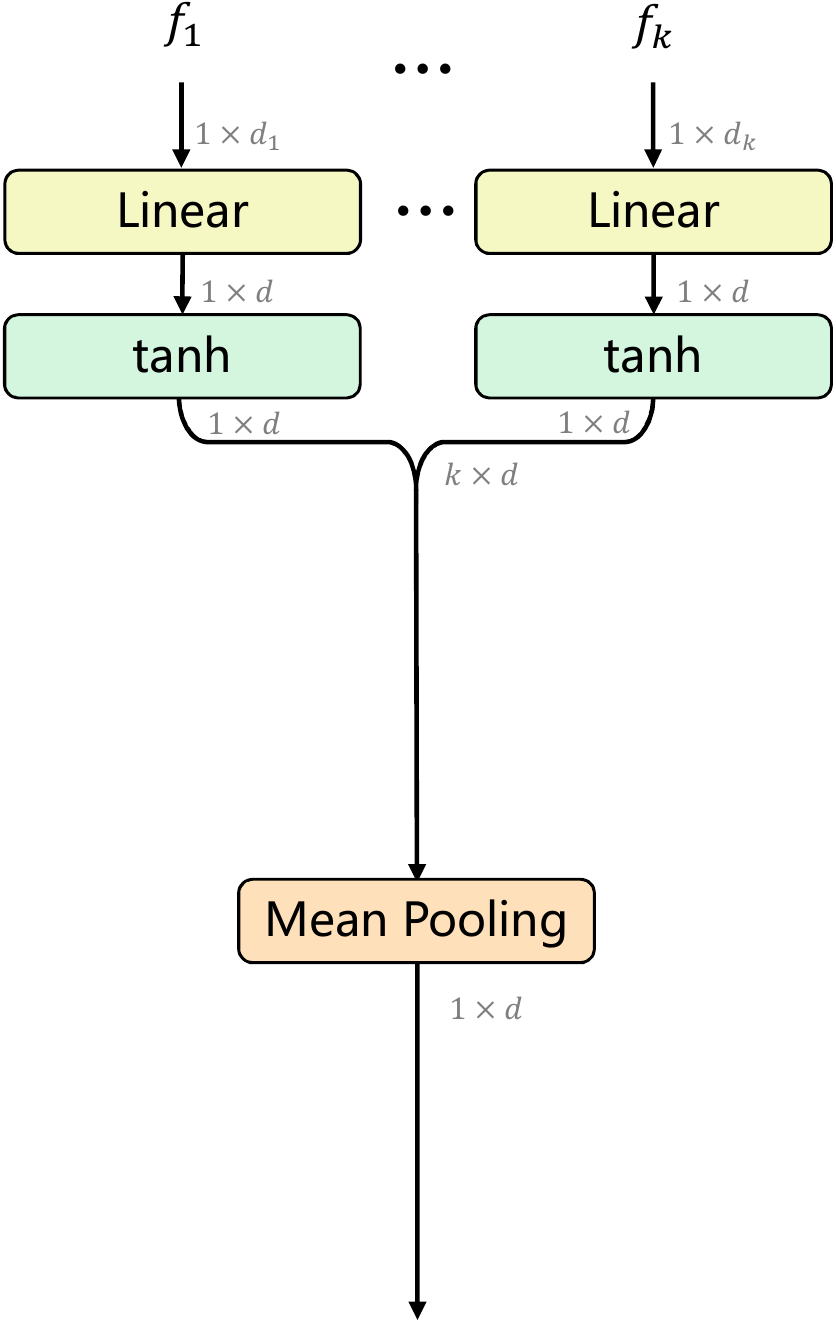}%
    \includegraphics[height=3cm]{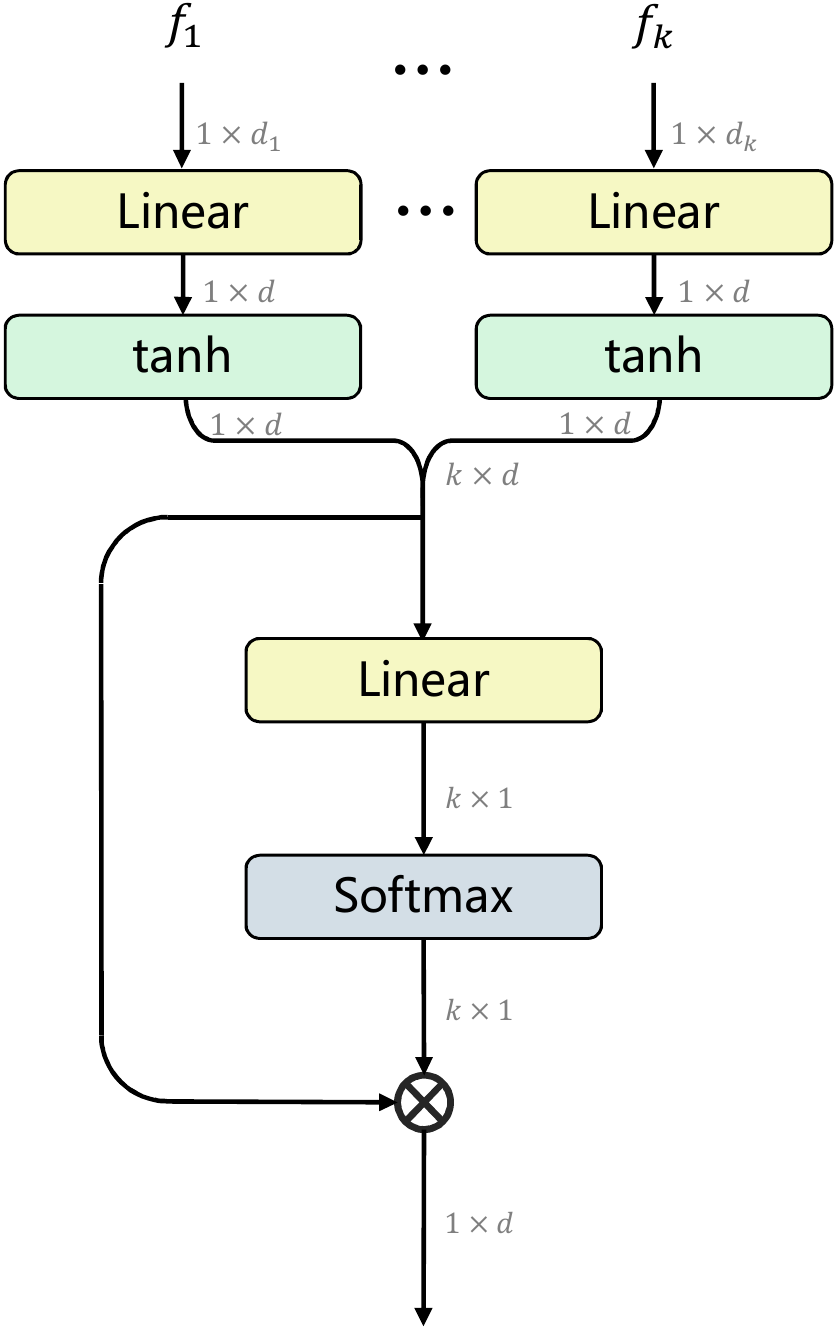}
  }%
}
\setlength{\twosubht}{\ht\twosubbox}


\centering

\subcaptionbox{\label{fig:mhsa} MHSA}{%
  \includegraphics[height=\twosubht]{fig/fig1a.pdf}%
}\hspace{-1pt}
\subcaptionbox{\label{fig:atten-free} Attention-free}{%
  \includegraphics[height=\twosubht]{fig/fig1b.pdf}%
}\hspace{8pt}
\subcaptionbox{\label{fig:laff} LAFF}{%
  \includegraphics[height=\twosubht]{fig/fig1c.pdf}%
}
\caption{\textbf{Three distinct blocks for feature fusion}: (a) Multi-Head Self-Attention (MHSA), (b) Attention-free, and (c) our proposed LAFF.}

\label{Figure:attention_block_vs_multi_head_attention}

\end{figure}

We propose in this paper a much simplified feature fusion block, termed Lightweight Attention Feature Fusion (LAFF), see \cref{fig:laff}. LAFF is generic, working for both video and text ends. Video/text features are combined in a convex manner in a specific LAFF block, with the combination weights learned to optimize cross-modal text-to-video matching. Performing fusion at the feature level, LAFF can thus be viewed as an early fusion method. Meanwhile, with the multi-head trick as used in MHSA, multiple LAFFs can be deployed within a single network, with their resultant similarities combined in a late fusion manner. The ability to perform feature fusion at both early and late stages and at both video and text ends makes LAFF a powerful method for exploiting diverse, \xredit{multi-level} (off-the-shelf) features for text-to-video retrieval. In sum, our main contributions are as follows: \\
$\bullet$ We are the first to study both video-end and text-end feature fusion for text-to-video retrieval. Given the increasing availability of deep vision/language models for feature extraction, this paper presents an effective mean to harness such dark knowledge for tackling the task. \\
$\bullet$ We propose LAFF, a lightweight feature fusion block, capable of performing fusion at both early and late stages. Compared to MHSA, LAFF is much more compact yet more effective. Its attentional weights can also be used for selecting fewer features,  with the retrieval performance mostly preserved. \\
$\bullet$ Experiments on five benchmarks, \ie MSR-VTT, MSVD, TGIF, VATEX and TRECVID AVS 2016-2020, show that the LAFF-based video retrieval model (\cref{Figure:overall_framework}) compares favorably against the state-of-the-art, resulting in a  strong baseline for text-to-video retrieval. Code is available at GitHub\footnote{\url{https://github.com/ruc-aimc-lab/laff}}.

\section{Related Work} \label{sec:related}

\textbf{Feature fusion for text-to-video retrieval}. Previous methods on feature fusion focus either  on the video end or on the text end, see \cref{tab:related_work}. For video-end feature fusion, earlier works often simply use vector concatenation to merge multiple features in advance to cross-modal representation learning \cite{dong2018predicting}.
JE \cite{mithun2018learning_r2} and its journal extension \cite{mithun2019joint_r3} \xredit{have made an initial attempt to combine diverse video features by late fusion, where multiple feature-specific common spaces are learned. Multi-space similarities are either averaged \cite{mithun2018learning_r2} or linearly combined with empirical weights \cite{mithun2019joint_r3}}.
CE~\cite{liu2019use} also uses late fusion, but resorts to a learning based method, \ie Mixture of Experts (MoE), to determine the fusion weights on the fly. 
MMT~\cite{gabeur2020multi} improves over CE by \xredit{first using} a multi-modal Transformer to aggregate features from the frame level to the video level, and later using MoE for combining multi-space similarities. JE, CE and MMT all use a single text feature,  so these methods leave text-end feature fusion untouched.

As for text-end feature fusion, W2VV++~\cite{LiXirong2019W2VVPP} takes an early fusion approach, combining the output of three text encoders, \ie bag-of-words, word2vec and GRU, by vector concatenation. By contrast, SEA~\cite{LiXirong2020SEA} opts for late fusion, first building a common space per text feature and then averaging the similarities computed within the individual spaces. The more recent TEACHTEXT  \cite{croitoru2021teachtext} first trains multiple CEs with different text encoders, and later combines these models by late fusion with MoE-predicted weights.

\begin{table}[tbh!]
\normalsize
\renewcommand\arraystretch{1.1}
\centering
\begin{center}
\caption{\textbf{Taxonomy of feature fusion methods for text-to-video retrieval}. Note that feature fusion is conceptually different from multi-level feature learning, \eg JPoSE~\cite{wray2019fine_r1},  PIE-Net~\cite{song2019polysemous_r4} and Dual Encoding \cite{Dong2021DE_hybrid}, where new features are first computed at varied levels from a single feature input and combined later. So research in that line is excluded.}
\label{tab:related_work}
\scalebox{0.68}{
\setlength{\tabcolsep}{5mm}{

\begin{tabular}{@{}llll@{}}
\toprule
\textbf{Method} & \textbf{Feature modality} & \textbf{Fusion stage} & \textbf{Fusion block} \\ \midrule
JE, ICMR18~\cite{mithun2018learning_r2} & Video & Late fusion & Average \\
JE, IJMIR19~\cite{mithun2019joint_r3} & Video & Late fusion & Manual weights \\
CE, BMVC19~\cite{liu2019use} & Video & Late fusion & MoE \\
MMT, ECCV20~\cite{gabeur2020multi} & Video & Hybrid fusion & Multimodal Transformer + MoE \\
W2VV++, MM19~\cite{LiXirong2019W2VVPP} & Text & Early fusion & Vector concatenation \\
SEA, TMM21~\cite{LiXirong2020SEA} & Text & Late fusion & Average \\
TEACHTEXT, ICCV21~\cite{croitoru2021teachtext} & Text & Late fusion & MoE \\
\textit{This work} & Text/Video & Hybrid fusion & LAFF 
\\ \bottomrule
\end{tabular}

}
}
\end{center}

\end{table}

\textbf{Attentional feature fusion in other contexts}.  
LAFF is conceptually different from context gating modules \cite{dauphin2017language, miech2018learning}, which aim to re-weight each dimension of a given feature vector, and thus produce weight per dimension. By contrast, as LAFF is to combine multiple
features, it produces weight per feature vector.
LAFF is technically inspired by attention-based multiple instance learning (MIL) \cite{icml18-mil}, wherein there is a need of aggregating multiple instance-level features into a case-level feature. However, the text-to-video retrieval task differs from MIL in the following two aspects, making the attention-based MIL not directly applicable in the new context. First, instances in a MIL setting are of the same modality, \eg patches taken from the same image \cite{icml18-mil} or images from an image array \cite{icpr20-rop}, so the instance-level features are homogeneous and directly comparable. By contrast, the video or text features to be fused are obtained by distinct feature extractors with varied feature dimensions and are thus incompatible. Second, MIL is typically exploited in the context of a classification task, so the case-level feature is used as input to a classification layer. By contrast, our fused  features are meant for cross-modal matching. Hence, a feature fusion block at one end shall be used in pair with a fusion block at the other end.
The technical novelty of LAFF lies in its overall design that effectively answers the unique challenges in video/text feature fusion for text-to-video retrieval. While combining homogeneous features
have been studied in other contexts \cite{MMM_LiCHYCX19,LiuCZLR20,eccv_WooPLK18}, our paper fills the gap
between diverse feature fusion and text-to-video retrieval.

\section{A New Baseline} \label{sec:method}

We propose trainable feature fusion for both video and text ends. Specifically, suppose we have a specific video $x$ represented by a set of $k_1$ video-level features, $\{f_{v,1}(x), \ldots, f_{v,k_1}(x)\}$, and a specific textual query $q$ represented by a set of $k_2$ sentence-level features $\{f_{t,1}(q), \ldots, f_{t,k_2}(q)\}$. We shall construct two feature fusion blocks to encode respectively the video and the query into their $d$-dimensional cross-modal embeddings $e(x)$ and $e(q)$. Their semantic similarity $s(x,q)$ is measured in terms of the two embeddings accordingly, \ie 
\begin{equation}
\left\{ 
\begin{array}{ll}
 e(x) &:= fusion_v(\{f_{v,1}(x), \ldots, f_{v,k_1}(x)\}),\\
  e(q) &:= fusion_t(\{f_{t,1}(q), \ldots, f_{t,k_2}(q)\}), \\
 s(x,q) & := similarity(e(x), e(q)).
\end{array} \right.
\label{eq:xdef}
\end{equation}
As such, text-to-video retrieval for the given query $q$ is achieved by sorting all videos in a test collection in light of their $s (x,q)$ in descending order. In what follows, we describe the proposed  LAFF as a unified implementation of the \emph{fusion} blocks in \cref{eq:xdef}, followed by its detailed usage for text-to-video retrieval.

\begin{figure}[htbp!]
\centering 
\includegraphics[width=1.0\textwidth]{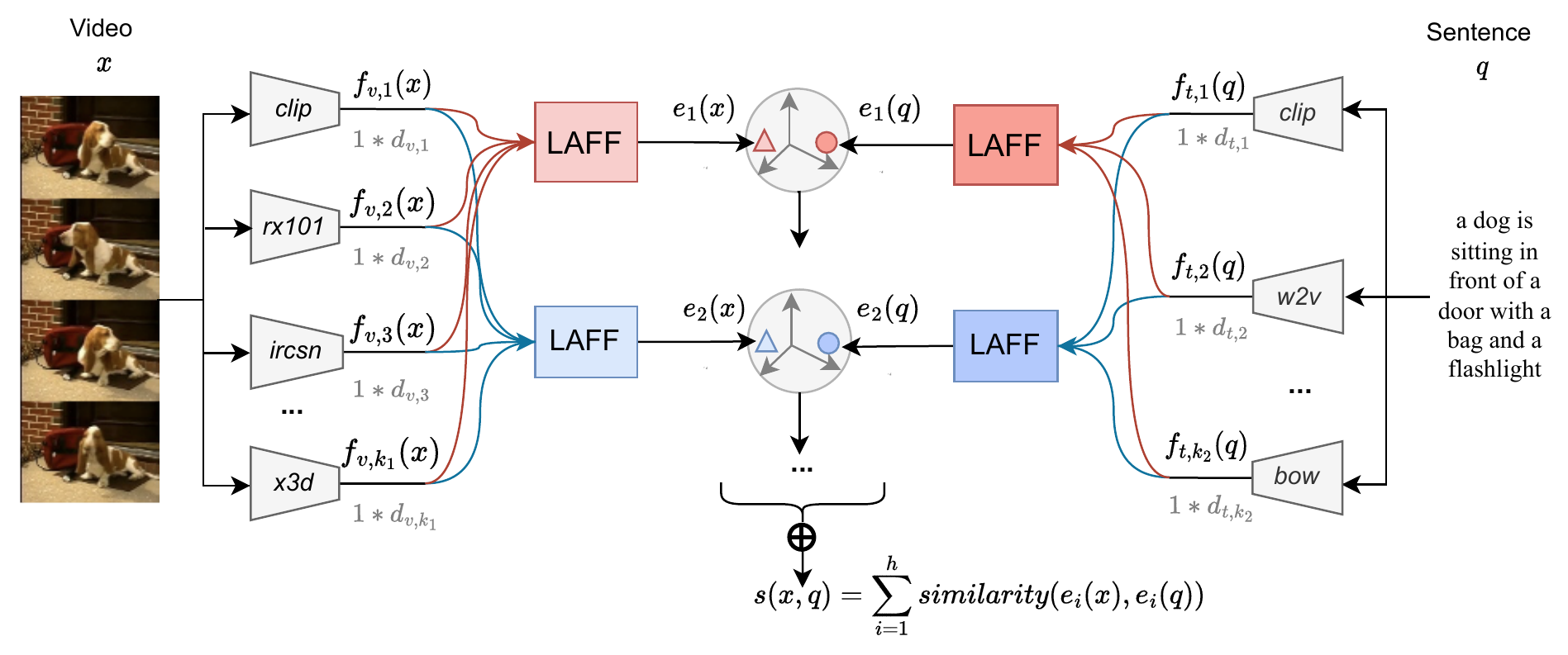} \caption{\textbf{Conceptual diagram of using paired LAFF blocks for text-to-video retrieval}. Given a specific video $x$, we employ multiple (pre-trained) feature extractors to obtain a set of $k_1$ video-level features $\{f_{v,1}(x),\ldots,f_{v,k_1}(x)\}$. In a similar manner we extract from a query sentence $q$ a set of $k_2$ sentence-level features  $\{f_{t,1}(q),\ldots,f_{t,k_2}(q)\}$. Each pair of LAFFs determines a common space. In particular, a specific pair of LAFFs, indexed by $i$, aggregates the video (sentence) features into a $d$-dimensional cross-modal feature $e_i(x)$ ($e_i(q)$), and consequently computes the video-text similarity $s_i(x,q)$ per space. The sum of the similarities from all $h$ common spaces, \ie $\sum_{i=1}^h s_i(x,q)$, is used for retrieval.} 
\label{Figure:overall_framework} 
\end{figure}

\subsection{The LAFF Block}


Without loss of generality, we are provided with a diverse set of $k$ different features $\{f_1, \ldots, f_k\}$, sized as $d_1, \ldots, d_k$, respectively. As the features are obtained by distinct extractors and thus incompatible, we shall use a feature transformation layer to rectify the diverse features to be of the same length. To convert the $i$-th feature to a new $d$-dimensional feature, we use 
\begin{equation} \label{eq:transform}
f'_i = \sigma (Linear_{d_i \times d} (f_i)),
\end{equation}
where $\sigma$ is a nonlinear activation function. As the output of the non-linear activation in LAFF is to calculate the cosine similarity, we use \emph{tanh} in this work\footnote{Other non-linear activations such as ReLU and sigmoid will make each dimension non-negative, constrain the feature space be in the first quadrant, and consequently put a lower boundary of $0$ on the cosine similarity. As such, the similarity will be less discriminative than the \emph{tanh} counterpart.}. The notation of $Linear_{d_i \times d}$ indicates a fully connected layer with an input of size $d_i$ and an output of size of $d$. Each input feature has its own $Linear$, optional when $d_i$ equals to $d$.

Although the transformed features $\{f'_i\}$ are now comparable, they are not equally important for representing the video/text content. We thus consider a weighted fusion, \ie
\begin{equation} \label{eq:combine_feats}
\bar{f} = \sum_{i}^k a_i f '_i,
\end{equation}
with weights $\{a_1, \ldots, a_k\}$ computed by a lightweight attentional layer as follow,
\begin{equation}\label{eq:light-att}
\{a_1, \ldots, a_k\} = softmax(Linear_{d\times1}(\{f'_1, \ldots, f'_k\})).
\end{equation}

As shown in \cref{fig:atten-free}, the Attention-free feature fusion block is a special case of LAFF when enforcing the weights in \cref{eq:combine_feats} to be uniform, \ie $a_i=\frac{1}{k}$. Compared to Attention-free, LAFF has $d$ more parameters to learn, see \cref{tab:params}. Such a small amount of extra parameters turn out to be important for improving the effectiveness of feature fusion, as our ablation study will show. Compared with MHSA, LAFF has much fewer trainable parameters and is thus more data-efficient. Furthermore, as the attentional weights of LAFF are directly used for a convex combination of the features, LAFF is more interpretable than MHSA.

\setlength{\tabcolsep}{4pt}
\begin{table}[tbh!]
\begin{minipage}[b]{1\linewidth} %
\begin{center}

\caption{\textbf{Complexity analysis of feature fusion blocks}, with $D$ indicating the overall dimension of the input features and $d$ as the dimension of the output features. FLOPs are computed based on input features of shape $8 \times 2048$. }
\label{tab:params}
\scalebox{1}{

\begin{tabular}{llr} 
\toprule
\textbf{Feature fusion block} & \textbf{Parameters} & \textbf{FLOPs (M)}	 \\ 
\midrule
MHSA  & $D\times d+4 \times d^2$ & 94.90 \\ 
Attention-free & $D\times d$ & 27.78 \\
LAFF & $D\times d+ d$ & 27.80 \\
\bottomrule
\end{tabular}

}
\end{center}
\end{minipage}%

\end{table}

\subsection{Paired LAFFs for Text-to-Video Retrieval}

\subsubsection{Network Architecture}

We now detail the usage of LAFF for text-to-video retrieval. A straightforward solution is to substitute LAFF for the $fusion$ functions in \cref{eq:xdef}. As such, we have a single configuration of how the video/text features are combined. However, due to the high complexity of the video and text contents, we hypothesize that the single configuration is suboptimal for cross-modal representation and matching. Borrowing the multi-head idea of MHSA, we consider multi-head LAFF. In particular, we deploy $h$ pairs of LAFFs, where each pair of LAFFs jointly determine a latent common space for video-text matching. In particular, a specific pair of LAFFs, denoted as $<LAFF_{v,i}, LAFF_{t,i}>$, aggregates the video/text features into a $d$-dimensional cross-modal embedding vector $e_i(x)$/$e_i(q)$, \ie 
\begin{equation} \label{eq:paired_laffs}
\left\{ \begin{array}{rl}
 e_i(x) &= LAFF_{v,i}(x) \\
e_i(q) &= LAFF_{t,i}(q) \\
s_i(x,q) &= similarity(e_i(x), e_i(q))
       \end{array} \right.
\end{equation}
where $similarity$ is the widely used cosine similarity. Accordingly, we compute the final video-text similarity as the mean of the $h$ individual similarities, 
\begin{equation}
    s(x,q) = \frac{1}{h}\sum_{i=1}^h s_i(x,q).
\end{equation}

The overall architecture is illustrated in Fig. \ref{Figure:overall_framework}. In order to make the amount of trainable parameters invariant with respect to $h$, we set $d=\frac{d_0}{h}$, where $d_0$ is a constant empirically set to 2,048. As such, the multi-head version of LAFF is not an ensemble. We use $h=8$, unless otherwise stated.

\begin{figure}[tbh!] 
\centering 
\includegraphics[width=0.7\textwidth]{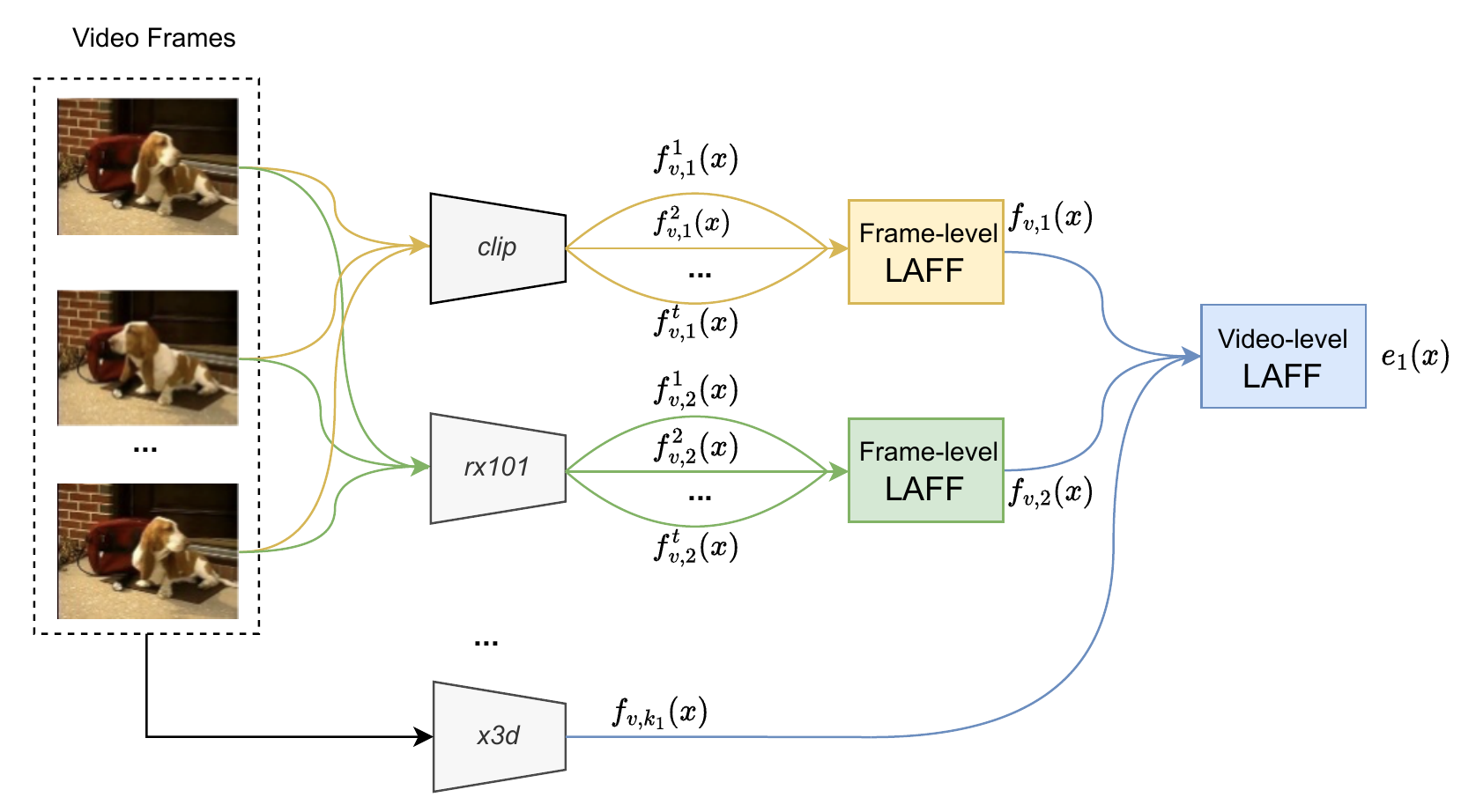} \caption{\textbf{LAFF-ml for multi-level feature fusion}. Frame-level LAFF is applied per feature, \eg \emph{clip} or \emph{rx101}. The outputs of the frame-level LAFF blocks are later combined (with other video-level features, \eg \emph{x3d}) by a video-level LAFF.} 
\label{fig:LAFF_multi_level} 
\end{figure}

\textbf{LAFF for multi-level feature fusion}. So far we presume the features to be fused are already at the video level. In fact, for its high flexibility, LAFF can be extended with ease to a multi-level variant to deal with the situation wherein different frame-level and video-level features coexist. Fig. \ref{fig:LAFF_multi_level} shows this variant, which we term \emph{LAFF-ml}. LAFF-ml works in a bottom-up manner, where a set of specific frame-level features are aggregated via a specific LAFF block to produce a video-level feature. Suppose there are two different frame-level features, \eg \emph{clip} and \emph{rx101}.  Each will have its own LAFF block. The (resultant) different video features are then fused via a video-level LAFF block.

\subsubsection{Network Training}

Following the good practice of the previous work, we adopt as our base loss function the triplet ranking loss with hard-negative mining \cite{bmvc_FaghriFKF18}. For a specific sentence $q$ in a given training batch, let $x_+$ and $x_-$ be videos relevant and irrelevant \wrt $q$, and $x^*_-$ be the hard negative that violates the ranking constraint the most. We have 
\begin{equation} \label{eq:base-loss}
\left\{ 
\begin{array}{ll}
x^*_- &=\operatorname{argmax}_{x_-} (s(x_-,q) - s(x_+,q)) \\ 
\operatorname{loss}(q)  &=\max (0, \alpha+s(x^*_-,q) - s(x_+, q)),
\end{array} \right.
\end{equation}
where $\alpha$ is a positive hyper-parameter controlling the margin of the ranking loss. 

As \cite{LiXirong2020SEA} has documented, when training a cross-modal network that produces multiple similarities, combining losses per similarity gives better results than using a single loss with the combined similarity. Hence, we follow this strategy, computing $\operatorname{loss}_i(q)$, namely the loss in the $i$-th space by substituting  $s_i$ for  $s$ in \cref{eq:base-loss}. The network is trained to minimize a combined loss $\sum_{i=1}^h \operatorname{loss}_i(q)$.

\section{Experiments} \label{sec:experimet}

In order to evaluate the effectiveness of LAFF, we conduct a series of experiments. An ablation study is performed on MSR-VTT, a \emph{de facto} benchmark, to evaluate LAFF in multiple aspects. We then compare our LAFF-based retrieval model with the state-of-the-art on MSR-VTT and three other popular benchmarks including MSVD, TGIF and VATEX. In order to assess our retrieval method on a much larger collection,  a post-competition evaluation is conducted on the TRECVID AVS benchmark series. 

\subsection{Common Setups}

\textbf{Implementation Details}. Eight video features and five text features are used, see \cref{tab:feature_intro}.
The margin $\alpha$ in the loss is set to 0.2 according to VSE++~\cite{bmvc_FaghriFKF18}. We perform SGD based training, with a mini-batch size of 128 and RMSProp as the optimizer. The learning rate is initially set to $10^{-4}$, decayed by a factor of 0.99 per epoch. Following \cite{JoulinMJV16}, we half the learning rate if the validation performance does not increase in three consecutive epochs. Early stop occurs when no validation performance increase is achieved in ten consecutive epochs. The dropout rate of the \emph{Linear} layers is set to 0.2. All experiments were done with PyTorch (1.7.1) \cite{PaszkePytorch19} on an Nvidia GEFORCE GTX 2080Ti GPU. 

\textbf{Evaluation Criteria}.
We report three standard rank-based metrics: Recall at Rank N (R@N, N=1, 5, 10), Median rank (Med r), and mean Average Precision (mAP) for assessing the overall ranking quality.

\newcommand{\tabincell}[2]{\begin{tabular}{@{}#1@{}}#2\end{tabular}}

\begin{table}[tbh!]
\caption{\textbf{Video/text features used in ablation study}. Video-level features are obtained by mean pooling over frames or segments, unless otherwise stated.}
\setlength{\abovecaptionskip}{1.5pt}
\resizebox{1\columnwidth}{!}{
\scriptsize
\centering

\label{tab:feature_intro}
\begin{tabular}{lrl}
    \toprule
    \textbf{Feature} &  \tabincell{l}{\textbf{Dim.}}  & \textbf{Short description }       \\ 
    \noalign{\smallskip}
    \hline
    \noalign{\smallskip}
    \multicolumn{3}{@{}l}{\textbf{\textit{Video features:}}} \\ 
    \textit{rx101}  & 2,048& ResNeXt-101 trained on the full set of ImageNet \cite{tomm20-shuffle}.  \\ 
    \textit{re152}  & 2,048 & ResNet-152  from the MXNet model zoo.   \\ 
    \textit{wsl}  & 2,048 & \tabincell{l}{ResNeXt-101  pre-trained by weakly supervised learning on 940 million public images, \\followed by fine-tuning on ImageNet1k \cite{wsl2018}.}     \\
    \textit{clip }  & 512  & CLIP (ViT-B/32) pre-trained on web images and corpus by contrastive learning \cite{2021clip_icml}. \\ 
    \textit{c3d}    & 2,048 & C3D  trained on Kinetics400 \cite{c3d2015}. \\ 
    \textit{ircsn} & 2,048 & irCSN-152  which trained by weakly supervised learning on IG-65M \cite{ircsn}.   \\  
    \textit{tf}     & 768  & TimeSformer trained on HowTo100M~\cite{timesformer}.  \\ 
    \textit{x3d}     & 2,048 & X3D trained on Kinetics400 \cite{feichtenhofer2020x3d}. \\
    \midrule
    \multicolumn{3}{@{}l}{\textbf{\textit{Text features:}}} \\ 
    \textit{bow}  & $m$    & \tabincell{l}{$m$-dimensional Bag-of-words feature, with $m$ being 7,675 (MSR-VTT), 2,916 (MSVD),\\ 3,980 (TGIF), or 10,312 (VATEX)} \\ 
    \textit{w2v}     & 500  & Word2Vec trained on Flickr tags \cite{dong2018predicting}.           \\ 
    \textit{gru}     & 1,024 & Mean pooling over hidden vectors of GRU trained from scratch  \cite{dong2018predicting}.   \\ 
    \textit{bert}     & 768  & The base version of BERT, pre-trained on BooksCorpus and English Wikipedia \cite{Devlin2019bert}. \\ 
    \textit{clip}    & 512  & The same CLIP as used to extract video features \\ \bottomrule 
    \end{tabular}
}
\end{table}

\subsection{Ablation Study} \label{ssec:eval-abla}



Our ablation study is conducted on MSR-VTT \cite{xu2016msr}, which has 10k videos in total, each associated with 20 captions. We adopt the official data split: 6,513 videos for training, 497 videos for validation and the remaining 2,990 images for test. In order to distinguish this data split from other customized splits, \eg JSFusion~\cite{yu2018joint}, we term the split \xredit{\textbf{MV-test3k}}.

\textbf{On Combining Diverse Video/Text Features}.
We investigate how LAFF responds when diverse video/text features are gradually added. For the ease of lateral comparison, we include as baselines the following two models: W2VV++ \cite{LiXirong2019W2VVPP}, which simply uses vector concatenation, and SEA \cite{LiXirong2020SEA} which learns cross-modal similarities per text feature.

\begin{figure}[htp]
\sbox\twosubbox{%
  \resizebox{\dimexpr\textwidth-1em}{!}{%
    \includegraphics[height=3cm]{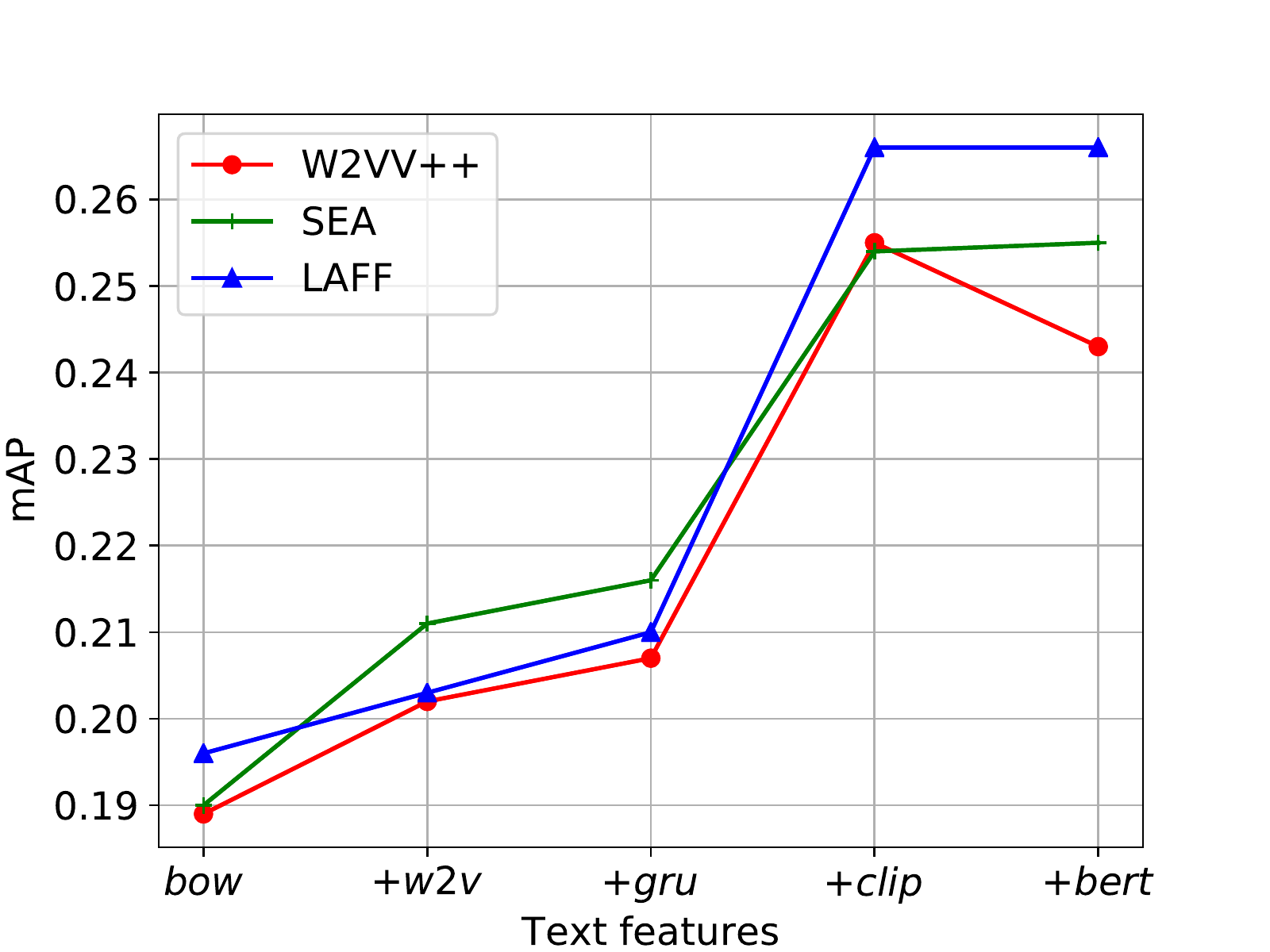} 
    \includegraphics[height=3cm]{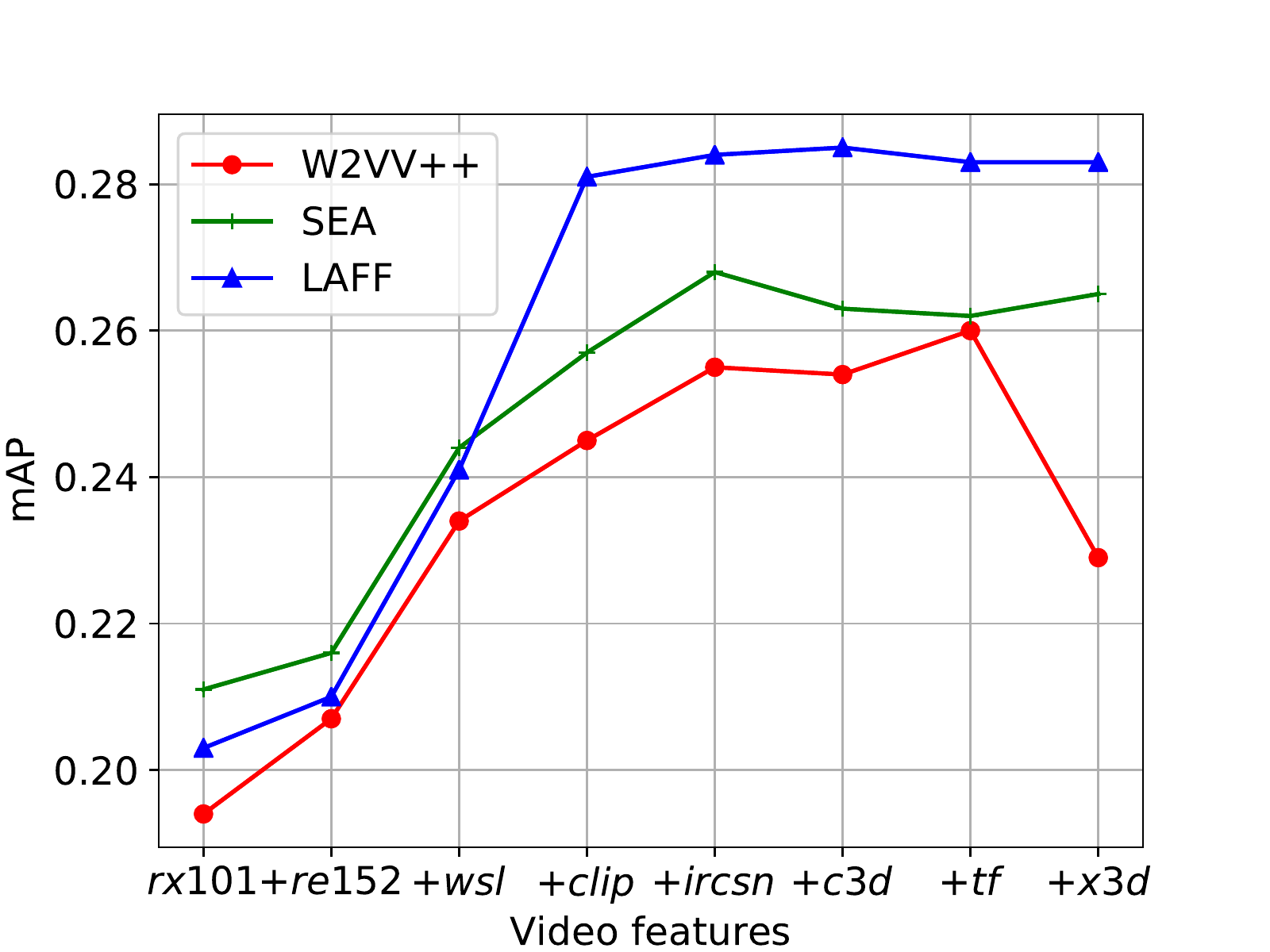}%
  }%
}
\setlength{\twosubht}{\ht\twosubbox}

\centering

\subcaptionbox{\label{fig:model_adjust_txt_encoder} Text feature fusion}{%
  \includegraphics[height=\twosubht]{fig/model_adjust_txt_encoder.pdf}%
}\quad
\subcaptionbox{\label{fig:model_adjust_vis_encoder} Video feature fusion}{%
  \includegraphics[height=\twosubht]{fig/model_adjust_vis_encoder.pdf}%
}
\caption{\textbf{Performance curves of three distinct models, \ie W2VV++, SEA and LAFF}, \wrt (a) text feature fusion, with \{\emph{rx101},\emph{re152}\} as video features, and (b) video feature fusion, with   \{\textit{bow,w2v,gru}\} as text features. LAFF is both effective and stable for fusing diverse features. Data: MV-test3k.}
\label{Figure:model_adjust_txt_vis_encoder}
\end{figure}

Given the many video and text features investigated in this work, a complete enumeration of video-text feature combinations is impractical. We choose to reduce the computation by only varying the features at one end, with features at the other end fixed.   \cref{fig:model_adjust_txt_encoder} shows the performance curves of W2VV++, SEA and LAFF \wrt text features, with \{\emph{rx101}, \emph{re152}\} as their common video features. The performance of all three models improves at the earlier steps when few features are fused. There is a noticeable drop in the performance curve of W2VV++ when \emph{bert} is included. LAFF is more effective and more stable. Similar results can be observed from  \ref{fig:model_adjust_vis_encoder}, which shows the performance curves of the three models \wrt video features. The above results justify the effectiveness of LAFF for combining diverse video/text features.

\textbf{Comparing Feature Fusion Blocks}. 
We compare the three feature fusion blocks by replacing LAFF in Fig. \ref{Figure:overall_framework} with MHSA and Attention-free, respectively. For a more fair comparison, we also apply the multi-loss trick on MHSA by optimizing  losses for different heads, denoted as MHSA(multi-loss). Moreover, we include as a baseline method that uses the simple feature concatenation strategy, as previously adopted in W2VV++ \cite{LiXirong2019W2VVPP}. The performance of text-to-video retrieval with specific feature fusion blocks is reported in  \cref{tab:four_method_with_fixed_video_feature}.
LAFF performs the best, followed by Attention-free, the concatenation baseline and MHSA. Attention-free, while being extremely simple, is more effective than MHSA for combining the increasing amounts of text features, with its mAP increases from 0.264, 0.321 to 0.326. The superior performance of LAFF against Attention-free (0.358 \emph{versus} 0.326) justifies the necessity of the attentional layer.

\begin{table}[tbh!]
\begin{center}
\caption{\textbf{Comparing feature fusion blocks}.
The simple feature concatenation used by W2VV++ is taken as a baseline. Numbers in parentheses are relative improvements against this baseline. Video features: all. Data: \xredit{MV-test3k}. }
\label{tab:four_method_with_fixed_video_feature}

\scalebox{0.75}{
\setlength{\tabcolsep}{3mm}{

\begin{tabular}{llrrrrl}
\toprule
 \textbf{Text features} & \textbf{Fusion block} & \multicolumn{1}{l}{\textbf{R1}} & \multicolumn{1}{l}{\textbf{R5}} & \multicolumn{1}{l}{\textbf{R10}} & \multicolumn{1}{l}{\textbf{Medr}} & \multicolumn{1}{l}{\textbf{mAP}} \\
\midrule
\multirow{5}*{\textit{bow, w2v, gru}} &
Baseline & 14.0 & 35.9 & 47.7 & 12 & 0.249 \\ 
&MHSA & 11.7 & 31.9 & 43.4 & 15 & 0.219 (12.0\%$\downarrow$) \\ 
& MHSA(mulit-loss) & 11.1 & 30.1 & 41.1 & 18 & 0.207  (16.8\%$\downarrow$) \\
&Attention-free & 15.4 & 37.8 & 49.7 & 11 & 0.264 (6.0\%$\uparrow$) \\ 
&LAFF & \textbf{16.0} & \textbf{39.5} & \textbf{51.4} & \textbf{10} & \textbf{0.276} (10.8\%$\uparrow$) \\ 
\hline
\noalign{\smallskip}
\multirow{5}*{\textit{bow, w2v, gru, clip}} &
Baseline & 19.2 & 43.5 & 55.3 & 8 & 0.310 \\ 
&MHSA & 18.8 & 43.0 & 54.6 & 8 & 0.305 (1.6\%$\downarrow$) \\ 
& MHSA(mulit-loss) & 18.7 & 43.1 & 54.5 & 8 & 0.305    (1.6\%$\downarrow$) \\
&Attention-free & 20.5 & 44.8 & 56.2 & 7 & 0.321 (3.5\%$\uparrow$) \\ 
&LAFF & \textbf{23.7} & \textbf{49.1} & \textbf{60.6} & \textbf{6} & \textbf{0.358} (15.5\%$\uparrow$) \\
\hline
\noalign{\smallskip}

\multirow{5}*{\textit{bow, w2v, gru, clip, bert}} &
Baseline & 14.5 & 35.0 & 46.1 & 13 & 0.247 \\ 
&MHSA & 17.9 & 41.6 & 53.3 & 9 & 0.294 (19.0\%$\uparrow$) \\
 & MHSA (mulit-loss) & 19.0 & 43.4 & 54.9 & 8 & 0.306 (23.9\%$\uparrow$) \\
&Attention-free & 20.9 & 45.3 & 56.9 & 7 & 0.326 (32.0\%$\uparrow$) \\ 
&LAFF & \textbf{23.8} & \textbf{49.0} & \textbf{60.3} & \textbf{6} & \textbf{0.358} (44.9\%$\uparrow$) \\ 
\bottomrule
\end{tabular}

}
}
\end{center}

\end{table}

\begin{figure}[tbh!] 
\centering 
\includegraphics[width=1\textwidth]{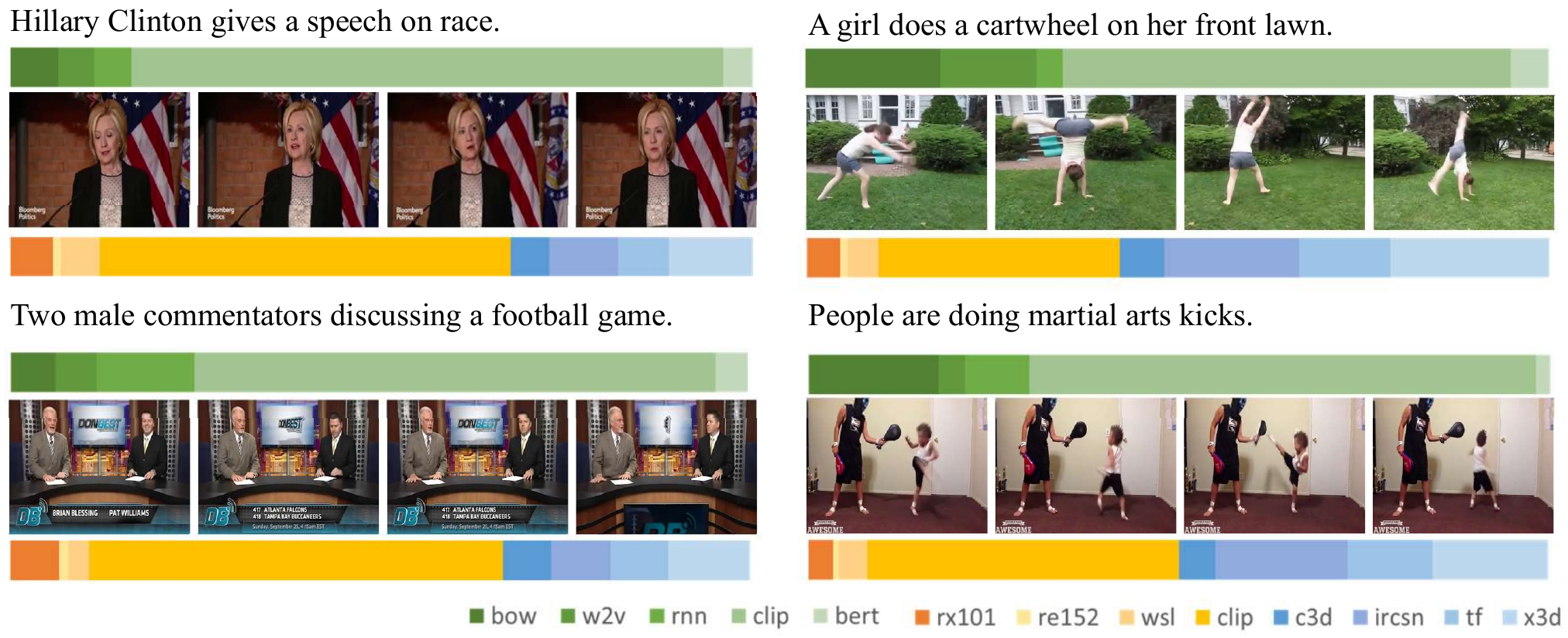} \caption{\textbf{Visualization of LAFF weights per feature}, with samples from the MV-test3k test set. Green, brown, and blue mean text features, 2D video features and 3D video features, respectively. Best viewed in color.} 
\label{Figure:weigth_exp} 
\end{figure}

\textbf{LAFF Weights for Model Interpretability and Feature Selection}.
\cref{Figure:weigth_exp} visualizes the LAFF weights of videos and their associated captions selected from the MV-test3k test set. We observe that 3D-CNN features receive more weight when the video content contains more motions, see \cref{Figure:weigth_exp}(b).  For each feature, its weight averaged over samples reflects its contribution to the retrieval performance. The weights of text features in descending order are \textit{clip} (64.3\%), \textit{bow} (15.7\%), \textit{gru}  (9.5\%), \textit{w2v} (6.5\%), \textit{bert} (4.0\%). For video feaetures, the order is  \textit{clip} (38.0\%), \textit{x3d} (16.8\%), \textit{ircsn} (13.3\%), \textit{tf} (10.9\%), \textit{rx101}(7.0\%), \textit{wsl}(6.6\%) , \textit{c3d} (5.1\%), \textit{re152} (1.4\%). We re-train our model with the top-3 ranked video / text features. Compared to the full setup (mAP of 0.358), the reduced model obtains mAP of 0.353, meaning a relatively small performance loss of 1.4\%. Hence, the LAFF weights are helpful for feature selection.

\textbf{Combined Loss \emph{versus} Single Loss}.
As \cref{tab:single_vs_multi_loss} shows, LAFF trained with the combined loss produces a relative  improvement of over 10\% in terms of mAP, when compared to its single-loss counterpart.

\begin{table}[tbh!]
\setlength{\abovecaptionskip}{0.cm}
\setlength{\belowcaptionskip}{-0.cm}

\normalsize
\renewcommand\arraystretch{1.0}

\begin{minipage}[]{0.48\linewidth}
\centering
\begin{center}
\caption{\textbf{Combined loss \emph{versus} single loss}. Video features: all.  }
\label{tab:single_vs_multi_loss}
\scalebox{0.5}{

\begin{tabular}{llrrrrl}
\toprule
\noalign{\smallskip}

\textbf{Text features} &\textbf{Loss} & \multicolumn{1}{l}{\textbf{R1}} & \multicolumn{1}{l}{\textbf{R5}} & \multicolumn{1}{l}{\textbf{R10}} & \multicolumn{1}{l}{\textbf{Med r}} & \multicolumn{1}{l}{\textbf{mAP}} \\ \midrule

\multirow{2}*{\textit{bow, w2v, gru}} &
Single & 14.1 & 36.2 & 47.8 & 12 & 0.250 \\ 
&Combined & \textbf{16.0} & \textbf{39.5} & \textbf{51.4} & \textbf{10} & \textbf{0.276} (10.4\%$\uparrow$) \\ 
\hline
\noalign{\smallskip}
\multirow{2}*{\textit{bow, w2v, gru, clip}} &
Single & 20.6 & 45.0 & 56.5 & 7 & 0.324 \\
&Combined & \textbf{23.7} & \textbf{49.1} & \textbf{60.6} & \textbf{6} & \textbf{0.358} (10.5\%$\uparrow$) \\ 
\bottomrule
\end{tabular}

}
\end{center}
\end{minipage} { }
\begin{minipage}[]{0.48\linewidth}
\begin{center}
\caption{\textbf{Effect of the number of common spaces} $h$.  Features: all. }
\label{tab:model_adjust_space}
\scalebox{0.5}{
\setlength{\tabcolsep}{5mm}{

\begin{tabular}{rrrrrr}
\toprule 
\multicolumn{1}{l}{\textbf{$h$}} & \multicolumn{1}{l}{\textbf{R1}} & \multicolumn{1}{l}{\textbf{R5}} & \multicolumn{1}{l}{\textbf{R10}} & \multicolumn{1}{l}{\textbf{Med r}} & \multicolumn{1}{l}{\textbf{mAP}} \\ \midrule \noalign{\smallskip}
1 & 23.1 & 48.3 & 59.9 & 6 & 0.352 \\
2 & 23.1 & 48.3 & 60.1 & 6 & 0.352 \\
4 & 23.5 & 49.1 & 60.4 & 6 & 0.356 \\
\textbf{8} & \textbf{23.7} & \textbf{49.1} & \textbf{60.6} & 6 & \textbf{0.358} \\
16 & 23.5 & 48.8 & 60.4 & 6 & 0.356 \\
\bottomrule \noalign{\smallskip}
\end{tabular}

}
}
\end{center}
\end{minipage}%

\end{table}
\setlength{\tabcolsep}{1.4pt}

\textbf{The Effect of the Number of Common Spaces}.
Concerning the number of common spaces $h$, we try different values, \ie \{1, 2, 4, 8, 16\}. As shown in \cref{tab:model_adjust_space}, the performance improves as $h$ increases, with the peak performance reached at $h=8$. 
While using a larger $h$ is beneficial, the relatively small gap between LAFF($h$=1) and LAFF($h$=8) suggests that the good performance of LAFF-based video retrieval is largely contributed by the LAFF block rather than the multi-space similarity.

To reveal how different are the embedding spaces to each other, we compute the Jaccard index between the top-5 video retrieval results of the individual spaces \wrt a specific query caption. The inter-space Jaccard index is lower than $0.5$, suggesting sufficient divergence. Nevertheless, whether videos/captions have been separated along different axes needs further investigation.

\subsection{Comparison with SOTA on Video Description Datasets} \label{ssec:sota_on_video_description_data}

\textbf{Datasets}. 
We further include MSVD \cite{chen2011collecting}, TGIF \cite{tgif2015} and VATEX \cite{wang2019vatex}. For MSVD and TGIF, we follow their official data splits. For VATEX, we follow the data split as used in HGR~\cite{chen2020fine}. As for MSR-VTT, in addition to the official MV-test3k split, we also report performance on another popular data split \cite{yu2018joint}, with 9k videos for training and 1k for test.  We term this split \xredit{\textit{MV-test1k}}.

\textbf{Baselines}.
We compare with the SOTA that uses the same data splits as we have mentioned. In particular, the following published models are included: JE \cite{mithun2019joint_r3}, 
W2VV++ \cite{LiXirong2019W2VVPP},
CE \cite{liu2019use}, 
TCE \cite{sigir20-yang-vr},
HGR \cite{chen2020fine},
SEA \cite{LiXirong2020SEA},
MMT \cite{gabeur2020multi},
DE \cite{Dong2021DE_hybrid},
SSB \cite{Patrick0AMHHV21},
SSML \cite{Amrani_Ben-Ari_Rotman_Bronstein_2021}, CLIP \cite{portillo2021straightforward},
CILP-FRL \cite{ChenHu2021}, and CLIP2Video (two-tower version\footnote{We prefer the two-tower version to the single-tower version, as the latter has to compute video and text embeddings online, making it not scalable for real applications.}) \cite{fang2021clip2video}.
In addition, we finetune CLIP per dataset, termed as CLIP-FT. \xredit{The video/text feature extracted by CLIP-FT is denoted as \emph{clip-ft}.}

Note that the video/text features used in the above models vary. For a head-to-head comparison, we re-train LAFF, LAFF-ml, JE, W2V++, SEA and MMT using the same set of selected features\footnote{Video features: \textit{clip-ft}, \textit{x3d}, \textit{ircsn} and \textit{tf}. Text features: \textit{clip-ft}, \textit{bow}, \textit{w2v} and \textit{gru}.}. 
Other baselines are not re-run, as they cannot handle the diverse video/text features without proper re-engineering. Since the fusion weights in JE have to be manually specified, we tried with three choices, \ie JE with uniform weights, JE (0.8 for \emph{clip-ft}) which assigns a much larger weight of 0.8 to the best \emph{clip-ft} feature (the remains features have equal weights) and JE (0.9 for \emph{clip-ft}).
When comparing with CLIP2Video, we let LAFF include the global video/text features extracted by CLIP2Video to see whether LAFF can flexibly harness new and more powerful features.

\textbf{Results}. 
 The performance of the different models on the multiple benchmarks is summarized in \cref{tab:compare_all}. Note that due to the inclusion of the better \emph{clip-ft} feature, the performance is better than that reported in the ablation study. 
 The baselines (JE, W2VV++ , SEA and MMT) get even worse results than using a single feature (\emph{clip-ft}).  The result   suggests that one cannot take for granted that adding better features will yield better performance, and an intellectual design of feature fusion is needed.
 The proposed LAFF consistently performs the best on all the test sets. LAFF-ml outperforms LAFF, which shows that flexible use of LAFF in multiple levels can further improve performance.

\begin{table}[!ht]
\setlength{\abovecaptionskip}{0.cm}
\setlength{\belowcaptionskip}{-0.cm}

\normalsize
\renewcommand\arraystretch{1.1}
\centering
\caption{\textbf{Comparison with the state-of-the-art on four benchmark datasets}, \ie MSR-VTT (MV-test3k/MV-test1k), MSVD, TGIF and VATEX. Baselines which use video/text features different from ours and thus not directly comparable are provided in the supplement.}
\label{tab:compare_all}
\scalebox{0.7}{
\begin{tabular}{@{}lrrrlrrrlrrrlrrrlrrr@{}}
\toprule
\multicolumn{1}{l}{\multirow{2}{*}{\textbf{Model}}} & \multicolumn{3}{c}{\textbf{MV-test3k}} & \multicolumn{1}{l}{} & \multicolumn{3}{c}{\textbf{MV-test1k}} & \multicolumn{1}{l}{} & \multicolumn{3}{c}{\textbf{MSVD}} & \multicolumn{1}{l}{} & \multicolumn{3}{c}{\textbf{TGIF}} & \multicolumn{1}{l}{} & \multicolumn{3}{c}{\textbf{VATEX}} \\ \cmidrule{2-4} \cmidrule{6-8} \cmidrule{10-12} \cmidrule{14-16}  \cmidrule{18-20} 
\multicolumn{1}{l}{} & \multicolumn{1}{l}{\textit{R1}} & \multicolumn{1}{l}{\textit{R5}} & \multicolumn{1}{l}{\textit{R10}} & \multicolumn{1}{l}{} & \multicolumn{1}{l}{\textit{R1}} & \multicolumn{1}{l}{\textit{R5}} & \multicolumn{1}{l}{\textit{R10}} & \multicolumn{1}{l}{} & \multicolumn{1}{l}{\textit{R1}} & \multicolumn{1}{l}{\textit{R5}} & \multicolumn{1}{l}{\textit{R10}} & \multicolumn{1}{l}{} & \multicolumn{1}{l}{\textit{R1}} & \multicolumn{1}{l}{\textit{R5}} & \multicolumn{1}{l}{\textit{R10}} & \multicolumn{1}{l}{} & \multicolumn{1}{l}{\textit{R1}} & \multicolumn{1}{l}{\textit{R5}} & \multicolumn{1}{l}{\textit{R10}} \\ \midrule
CLIP-FT (\emph{this paper}) & 27.7 & 53.0 & 64.2 &  & 39.7 & 67.8 & 78.4 &  & 44.6 & 74.7 & 84.1 &  & 21.5 & 40.6 & 49.9 &  & 53.3 & 87.5 & 94.0 \\
\multicolumn{20}{@{}l}{\textbf{\textit{The same video and text feature as ours}}}  \\
JE \cite{mithun2019joint_r3} (uniform weights) & 21.2 & 46.5 & 58.4 &  & 36.0 & 65.9 & 76.4 &  & 35.9 & 71.0 & 81.8 &  & 18.7 & 37.5 & 47.1 &  & 50.2 & 88.7 & 95.4 \\
JE (0.8 for \emph{clip-ft}) & 26.1 & 51.7 & 63.3 &  & 41.2 & 73.2 & 82.5 &  & 39.4 & 69.9 & 79.4 &  & 21.7 & 41.3 & 50.9 &  & 54.1 & 89.0 & 95.0 \\
JE (0.9 for \emph{clip-ft}) & 25.9 & 51.4 & 63.0 &  & 40.9 & 72.7 & 82.1 &  & 38.8 & 69.7 & 78.9 &  & 21.3 & 40.9 & 50.3 &  & 53.5 & 88.3 & 94.6 \\
W2VV++ \cite{LiXirong2019W2VVPP} & 23.0 & 49.0 & 60.7 &  & 39.4 & 68.1 & 78.1 &  & 37.8 & 71.0 & 81.6 &  & 22.0 & 42.8 & 52.7 &  & 55.8 & 91.2 & 96.0 \\
SEA \cite{LiXirong2020SEA} & 19.9 & 44.3 & 56.5 &  & 37.2 & 67.1 & 78.3 &  & 34.5 & 68.8 & 80.5 &  & 16.4 & 33.6 & 42.5 &  & 52.4 & 90.2 & 95.9 \\
MMT \cite{gabeur2020multi} & 24.9 & 50.5 & 62.0 &  & 39.5 & 68.3 & 78.3 &  & 40.6 & 72.0 & 81.7 &  & 22.1 & 42.2 & 51.7 &  & 54.4 & 89.2 & 95.0 \\
LAFF & 28.0 & 53.8 & 64.9 &  & 42.2 & 70.7 & \textbf{81.2} &  & 45.2 & 75.8 & 84.3 &  & 24.1 & 44.7 & 54.3 &  & 57.7 & 91.3 & 95.9 \\
LAFF-ml & \textbf{29.1} & \textbf{54.9} & \textbf{65.8} &  & \textbf{42.6} & \textbf{71.8} & 81.0 &  & \textbf{45.4} & \textbf{76.0} & \textbf{84.6} &  & \textbf{24.5} & \textbf{45.0} & \textbf{54.5} &  & \textbf{59.1} & \textbf{91.7} & \textbf{96.3} \\ \midrule
\multicolumn{20}{@{}l}{\textbf{\textit{Comparison with arXiv SOTA}}}  \\
CLIP2Video \cite{fang2021clip2video} & n.a & n.a & n.a &  & 44.5 & 71.3 & 80.6 &  & 44.7 & 74.8 & 83.7 &  & n.a & n.a & n.a &  & 54.8 & 89.1 & 95.1 \\
LAFF & n.a & n.a & n.a &  & \textbf{45.8} & \textbf{71.5} & \textbf{82.0} &  & \textbf{45.4} & \textbf{75.5} & \textbf{84.1} &  & n.a & n.a & n.a &  & \textbf{58.3} & \textbf{91.7} & \textbf{96.3} \\ 
\bottomrule
\end{tabular}
}
\end{table}

\subsection{Comparison with SOTA on TRECVID AVS 2016-2020}

\textbf{Setup}. 
The test collection for TRECVID AVS 2016-2018 (TV16/TV17/TV18) is IACC.3~\cite{iacc2009} with 335,944 video clips. The test collection for TRECVID AVS 2019–2020 (TV19/TV20) is V3C1~\cite{V3C1Fabian2019} with 1,082,659 video clips. We use the official metric, \ie inferred Average Precision (infAP) \cite{yilmaz2006estimating}.

\textbf{Baselines}.  
Due to the prominent performance of CLIP-FT and CLIP2Video as shown in \cref{ssec:sota_on_video_description_data}, we again compare with the two models. Since the top-3 ranked solutions of the AVS evaluation naturally reflect the state-of-the-art, we  include them as well. CLIP2Video was trained on the MV-test1k split. So for a fair comparison, we train LAFF and CLIP-FT on the same split.

\begin{table}[h!]
\setlength{\abovecaptionskip}{0.cm}
\setlength{\belowcaptionskip}{-0.cm}

\normalsize
\renewcommand\arraystretch{1}
\centering

\begin{minipage}[]{1\linewidth}
\begin{center}
\caption{\textbf{State-of-the-art performance on TRECVID AVS 2016--2020.}}
\label{tab:TRECVID_result}
\scalebox{0.7}{
\setlength{\tabcolsep}{3mm}{

\begin{tabular}{@{}lllllll@{}}
\toprule
\textbf{Model} & \textbf{TV16} & \textbf{TV17} & \textbf{TV18} & \textbf{TV19} & \textbf{TV20} & \textbf{MEAN} \\ \midrule
Rank 1 & {0.054}~\cite{tv16-nii}  & {0.206}~\cite{tv17-uva} & {0.121}~\cite{tv18-rucmm} & {0.163}~\cite{tv19-alibaba} & {\textbf{0.354}~\cite{tv20-aim3}} & n.a \\ 
Rank 2 & {0.051}~\cite{tv16-certh} & {0.159}~\cite{tv17-waseda} & {0.087}~\cite{tv18-inf} & {0.160}~\cite{tv19-rucmm} & {0.269~\cite{tv20-lirenmin}} & n.a \\ 
Rank 3 & {0.040}~\cite{tv16-inf} & {0.120}~\cite{tv17-vireo} & {0.082}~\cite{tv18-ntu} & {0.123}~\cite{tv19-waseda} & {0.229~\cite{tv20-wuvireo}} & n.a \\ 
\midrule
CLIP2Video & {0.176} & {0.229} & {0.114} & {0.176} & {0.207} & {0.180} \\ 
CLIP-FT & {0.191} & {0.215} & {0.105} & {0.147} & {0.203} & {0.172} \\ 
LAFF & {0.211} & {0.285} & {0.137} & {\textbf{0.192}} & {0.265} & {\textbf{0.218}} \\ 

LAFF-ml & {\textbf{0.222}} & {\textbf{0.290}} & {\textbf{0.147}} & {0.181} & {0.245} & {0.217} \\ \bottomrule
\end{tabular}
}

}
\end{center}
\end{minipage} 

\end{table}

\textbf{Results}. As shown in \cref{tab:TRECVID_result}, LAFF performs the best on TV16--TV19. Note that the top performer of TV20 was trained on the joint set of MSR-VTT, TGIF and VATEX. Re-training LAFF on this larger dataset results in infAP of 0.358, marginally better than the top performer.
We also conduct a case study on TV20, which shows that LAFF outperforms the CLIP series for action related queries with a large margin, see supplementary materials.
We attribute this result to the fact that LAFF integrates 3D-CNN features (\textit{ircsn}, \textit{c3d} and \textit{tf}), which were designed to capture action and motion information in the video content.

\section{Conclusions} \label{sec:conclusion}

For video retrieval by text, we propose LAFF, an extremely simple feature fusion block. LAFF is more effective than Multi-head Self-Attention, yet with much fewer parameters. Moreover, the attentional weights produced by LAFF can be used to explain the contribution of the individual video/text features for cross-modal matching. Consequently, the weights can be used for feature selection for building a more compact video retrieval model. Our LAFF-based video retrieval model surpasses the state-of-the-art on MSR-VTT, MSVD, TGIF, VATEX and TRECVID AVS 2016-2020. Given the increasing availability of (deep) video/text  features, we believe our work opens up a promising avenue for further research. 

\medskip

\textbf{Acknowledgments}. This work was supported by NSFC (No. 62172420, No. 62072463), BJNSF (No. 4202033), and Public Computing Cloud, Renmin University of China.

\clearpage
%
%
\bibliographystyle{splncs04}
\bibliography{paper-4413}

\end{document}